\documentclass{aa}

\usepackage[varg]{txfonts}
\usepackage{natbib}
\usepackage{blindtext}
\usepackage{booktabs}
\usepackage{pifont}
\usepackage{lscape}
\usepackage{multirow}
\usepackage{tabularx}
\usepackage{pbox}
\usepackage{wasysym}
\usepackage{cleveref}
\usepackage{upgreek}
\usepackage{listings}
\usepackage{color}

\newcommand{\nh}{\ensuremath{N_\text{H}}\xspace}

\newcommand{\xmm}{\textit{XMM-Newton}\xspace}

\newcommand{\spitzer}{\textit{Spitzer}\xspace}
\newcommand{\chandra}{\textit{Chandra}\xspace}

\begin{document}

\title{Extended X-ray emission in PKS~1718$-$649}
\author{T.~Beuchert\inst{\ref{inst1},\ref{inst2},\ref{inst3}}
\and A.~Rodr{\'i}guez-Ardila\inst{\ref{inst4},\ref{inst5}}
\and V.A.~Moss\inst{\ref{inst6}}
\and R.~Schulz\inst{\ref{inst6},\ref{inst3}}
\and M.~Kadler\inst{\ref{inst3}}
\and J.~Wilms\inst{\ref{inst2}}
\and R.~Angioni\inst{\ref{inst7},\ref{inst3}}
\and J.R.~Callingham\inst{\ref{inst6}}
\and C.~Gr\"{a}fe\inst{\ref{inst2}}
\and F.~Krau{\ss}\inst{\ref{inst8}}
\and A.~Kreikenbohm\inst{\ref{inst3}}
\and M.~Langejahn\inst{\ref{inst3}}
\and K.~Leiter\inst{\ref{inst3}}
\and F.M.~Maccagni\inst{\ref{inst9}}
\and C.~M{\"u}ller\inst{\ref{inst10},\ref{inst7}}
\and R.~Ojha\inst{\ref{inst11}}
\and E.~Ros\inst{\ref{inst7},\ref{inst12},\ref{inst13}}
\and S.J.~Tingay\inst{\ref{inst14}}
}

\institute{Anton Pannekoek Institute for Astronomy, Universiteit van Amsterdam, Science Park 904, 1098XH, Amsterdam, The Netherlands
\label{inst1}
\and Dr. Remeis-Observatory \& Erlangen Centre for Astroparticle
  Physics, Universit\"at Erlangen-N\"urnberg, Sternwartstrasse 7,
  96049 Bamberg, Germany 
\label{inst2}
\and Lehrstuhl f\"ur Astronomie, Universit\"at W\"urzburg, 
Emil-Fischer-Straße 31, 97074, W\"urzburg, Germany
\label{inst3}
\and Laborat\'orio Nacional de Astrof\'isica/MCTIC, Rua dos Estados Unidos, 154, Bairro das Na\c{c}\~{n}es, Itajub\'a, MG, Brazil
\label{inst4}
\and Instituto de Astrof\'{i}sica de Canarias, C/V\'{i}a L\'{a}ctea, s/n, 38206 La Laguna, Tenerife, Spain
\label{inst5}
\and Netherlands Institute for Radio Astronomy (ASTRON), PO Bus 2, 7990AA Dwingeloo, The Netherlands
\label{inst6}
\and Max-Planck-Institut für Radioastronomie, Auf dem Hügel 69, 53121, Bonn, Germany
\label{inst7}
\and GRAPPA \& Anton Pannekoek Institute for Astronomy, Universiteit van Amsterdam, Science Park 904, 1098XH Amsterdam, The Netherlands
\label{inst8}
\and INAF-Osservatorio Astronomico di Cagliari, Via della Scienza 5, 09047 Selargius (CA), Italy
\label{inst9}
\and IMAPP/Department of Astrophysics, Radboud University, Postbus/P.O.box 9010, 6500GL Nijmegen, The Netherlands
\label{inst10}
\and NASA/GSFC, Mail Code: 661, Greenbelt, MD 20771, USA
\label{inst11}
\and Observatori Astron\`omic, Universitat de Val\`encia, Parc Cient\'{\i}fic, C. Catedr\'atico Jos\'e Beltr\'an 2, 46980 Paterna, Val\`encia, Spain
\label{inst12}
\and Departament d'Astronomia i Astrof\'{\i}sica, Universitat de Val\`encia, C. Dr. Moliner 50, 46100 Burjassot, Val\`encia, Spain
\label{inst13}
\and International Centre for Radio Astronomy Research (ICRAR), Curtin University, Bentley, WA 6102, Australia 
\label{inst14}
}

\date{21-03-2018 /
10-04-2018 }

\abstract {PKS 1718$-$649 is one of the closest and most
  comprehensively studied candidates of a young active galactic
  nucleus (AGN) that is still embedded in its optical host galaxy. The
  compact radio structure, with a maximal extent of a few parsecs,
  makes it a member of the group of compact symmetric objects
  (CSO). Its environment imposes a turnover of the radio synchrotron
  spectrum towards lower frequencies, also classifying PKS 1718$-$649
  as gigahertz-peaked radio spectrum (GPS) source. Its close proximity
  has allowed the first detection of extended X-ray emission in a
  GPS/CSO source with \chandra\ that is for the most part unrelated to
  nuclear feedback. However, not much is known about the nature of
  this emission. By co-adding all archival \chandra\ data and
  complementing these datasets with the large effective area of \xmm,
  we are able to study the detailed physics of the environment of PKS
  1718$-$649. Not only can we confirm that the bulk of the
  $\lesssim$kiloparsec-scale environment emits in the soft X-rays, but we
  also identify the emitting gas to form a hot, collisionally ionized
  medium. While the feedback of the central AGN still seems to be
  constrained to the inner few parsecs, we argue that supernovae are
  capable of producing the observed large-scale X-ray emission at a
  rate inferred from its estimated star formation rate.}


\maketitle

\section{Introduction}\label{sec:intro}
With a turnover at around 4\,GHz, PKS~1718$-$649 is a prominent
representative of the class of gigahertz-peaked radio spectrum (GPS)
sources \citep{Tingay1997,Jauncey1998,Tingay2003}. It is one of the
closest sources of its kind, with only NGC~1052 and PKS~2254$-$367
being closer.  Spectral variations of the radio continuum led
\citet{Tingay2015} to argue for variable free-free absorbing (FFA) and
ionized foreground material as opposed to synchrotron-self absorption
(SSA) of jet-intrinsic plasma. High-spatial-resolution Very Long
Baseline Interferometry (VLBI) observations at 22\,GHz
\citep{Tingay2003} and 8.4\,GHz \citep{Ojha2004,Ojha2010} consistently
confirm a compact double structure of $\sim$10\,mas diameter, that is,
$\sim$3\,pc at a luminosity distance of 64.3\,Mpc ($z\sim 0.014428 \pm
0.000023$; \citealt{Meyer2004}).  This compact radio morphology with
two distinct hot spots classifies PKS~1718$-$649 as a compact
symmetric object (CSO). \citet{Tingay2002} provide an upper limit on
the separation speed of $\lesssim$0.08\,c. Together with the compact
extent of the radio structure, this translates to an approximate age
of $\gtrsim$60\,yr. At an age of only hundreds to thousands of years,
CSOs, where small advance speeds of terminal hotspots have been
constrained \citep{Owsianik1998}, are therefore considered to be
young.  An extended jet (similar to NGC~1052) beyond the known
parsec-scale VLBI structure of PKS~1718$-$649 seems unlikely. The ATCA
data presented by \citet{Maccagni2014} are also consistent with a
point source. Given the uv-coverage of the Murchison Widefield Array
(MWA), we can spectrally exclude extended radio emission at hundreds
of megahertz down to arcsecond scales. In the image plane, however, we
are not sensitive to faint remnant radio structures on these scales --
a regime that the future observatories MeerKAT and SKA are able to
probe. Moreover, PKS~1718$-$649 is the first young radio galaxy
confirmed to be $\gammaup$-ray bright \citep{Migliori2016}.

The inverted radio spectrum below a few gigahertz and the morphology of
PKS~1718$-$649 shape the picture of an active galactic nucleus (AGN) embedded in a cocoon of
ionized matter. Recently,
\citet{Maccagni2014,Maccagni2016,Maccagni2018} also provided
evidence for circumnuclear and clumpy molecular matter that is feeding
the new-born AGN. While the line of sight is piercing this obscuring
matter, both an inverted radio and photo-absorbed X-ray continuum are
expected. Studying the X-ray emission alongside with the radio
emission is therefore a valuable tool for a better understanding of
the environment of this young AGN (see also
\citealt{Mueller2016,Mueller2015} for a radio and X-ray study of the
other $\gammaup$-ray loud young radio galaxy PMN~J1603$-$4904).
While we also provide measures for the X-ray continuum absorption, here we
primarily concentrate on the X-ray emission detected from the
environment of PKS~1718$-$649.

An exemplary CSO, where extended X-ray emission could be investigated
with the unprecedented spatial resolution of \chandra\ in great
detail, is NGC~1052 \citep{Kadler2004a,Boeck2012}. Here, a
double-sided radio jet reaches into the kiloparsec (kpc)-scale environment and is
observed to align with collisionally ionized X-ray-bright gas.
\citet{Siemiginowska2016} were the first to systematically study the
X-ray signatures in a large number of CSOs. Within their sample,
extended kiloparsec-scale X-ray emission could only be detected for the
nearby source PKS~1718$-$649. To date, the lack of sufficient count
statistics, however, has made it impossible to unveil the nature of
this X-ray emitting gas. In this letter, we present novel results from
a recent \xmm\ observation and a stack of three archival
\chandra\ datasets, which combine the large effective area of
\xmm\ with the imaging capabilities of \chandra.

We use the cosmological parameters
$\Omega_\mathrm{m}=0.308$, $\Omega_{\uplambda}=0.692$, and
$H_{0}=67.8\,\mathrm{km}\,\mathrm{s}^{-1}\,\mathrm{Mpc}^{-1}$ \citep{Planck2016} and find
a correspondence of $1\,\mathrm{arcsec}\sim 312\,\mathrm{pc}$.

\section{Observations and data reduction}
\label{sec:obs}
\begin{table}
  \caption{List of the X-ray observations used in this letter.}
\label{tab:obs}
  \centering
  \scriptsize
  \resizebox{\columnwidth}{!}{
  \begin{tabular}[ht]{llllllllllllll}
    \hline\hline 
    abbrv. & date & obsid     & det   & exp [ks] & cnts [$\times 10^{2}$]  \\
    \hline
    Ch~1 & 2010-11-09 & 12849 & ACIS-S & 4.8 & 3.1 \\
    Ch~2 & 2014-06-20 & 16070 & ACIS-S & 15.9 & 14.2 \\
    Ch~3 & 2014-06-23 & 16623 & ACIS-S & 33.0 & 35.3 \\
    XMM  & 2017-03-05 & 0784530201 & EPIC-pn & 20.3 & 50.7 \\
    \hline
  \end{tabular}
}
\end{table}
We base our analysis on four archival X-ray observations (three by
\chandra\ and one by \xmm, all listed in
Table~\ref{tab:obs}).

For all \chandra\ observations, the source is observed with the back-illuminated chip
S3 of the ACIS-S CCD \citep{Garmire2003}. The VFAINT mode was used to most effectively
screen cosmic ray events. We make use of \texttt{CIAO~v.4.8.1} and \texttt{CALDB~v.4.7.2}
to reprocess event files with the task \texttt{chandra\_repro} and extract spectra using
\texttt{specextract} from within regions of 14\arcsec\ radius around the source pointing
center. We extract the background spectra from annuli of
30\arcsec--44\arcsec\ radius. Pileup can be neglected after fitting an absorbed power-law
based on the pileup-kernel in \texttt{ISIS~v.1.6.2-40}. Data are only extracted between
0.5\,keV and 8\,keV with maximal effective area. We rebin each \chandra\ spectrum to 4, 6,
and 8 channels per bin within 0.5--1\,keV, 3--5\,keV, and 5--8\,keV, respectively. That
way the grid oversamples the spectral resolution by no more than a factor of three
\citep{Kaastra2016}. We exclude a point source at RA:17$^{\rm h}$23$^{\rm
  m}$42$^{\rm s}$, Dec:$-$65$^{\circ}$00$^{''}$23$^{'}$ and extract surface brightness
profiles using \texttt{dmextract} with 15 annuli of 1\arcsec\ width and
$0.5+n$\arcsec\ radius ($n=0\ldots 14$), each centered at the source center. The
background from an annulus between 30\arcsec\ and 60\arcsec\ is subtracted from the
profile. The \chandra\ PSF is simulated for the spectrum Ch~2 by combining 25 runs of
\texttt{ChaRT}. For each realization, the PSF is projected onto the detector plane using
\texttt{MARX}. The resulting angular resolution is limited by the detector pixel size of
$\sim$0.5\arcsec ($\sim$156\,pc) at the given distance of PKS~1718$-$649.

We observed PKS~1718$-$649 using \xmm/EPIC-pn \citep{Villa1996,Meidinger1996,Strueder2001}
in Large Window mode and extracted the count spectrum using \texttt{SAS~v.16.0.0}.  After
creating calibrated event lists with filtered hot and bad pixels, events in the range
7--15\,keV are screened for particle flaring with a threshold of
$8\,\mathrm{cnts}\,\mathrm{ks}^{-1}\,\mathrm{arcmin}^{-2}$. We extract source counts from
a circular region of 40\arcsec\ radius for EPIC-pn and background counts from an off-source
region of 49\arcsec\ radius. The task \texttt{epatplot} returns no signs for pileup and we
consider all counts between 0.3 and 10\,keV. Following the same strategy as for \chandra,
we apply a geometrical binning with factors of 5, 6, 10 and 20 in the ranges 0.3--1.5\,keV,
1.5--2\,keV, 2--7\,keV, and 7--10\,keV, respectively.

\section{X-Ray image and spectral analysis}
\label{sec:spectra}
\begin{figure*}
  \centering
  \includegraphics[height=0.36\textwidth]{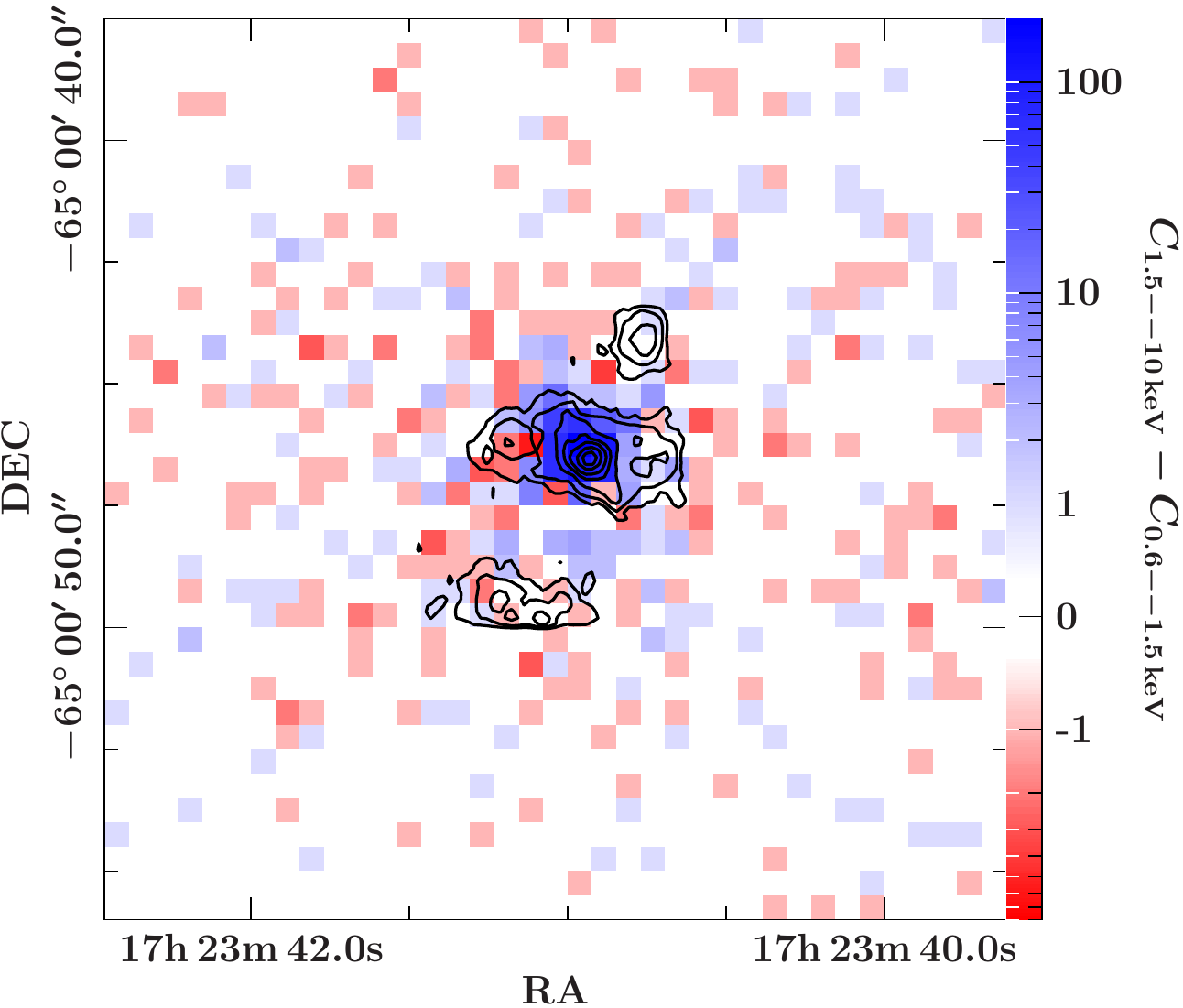}\hfill
  \includegraphics[height=0.36\textwidth]{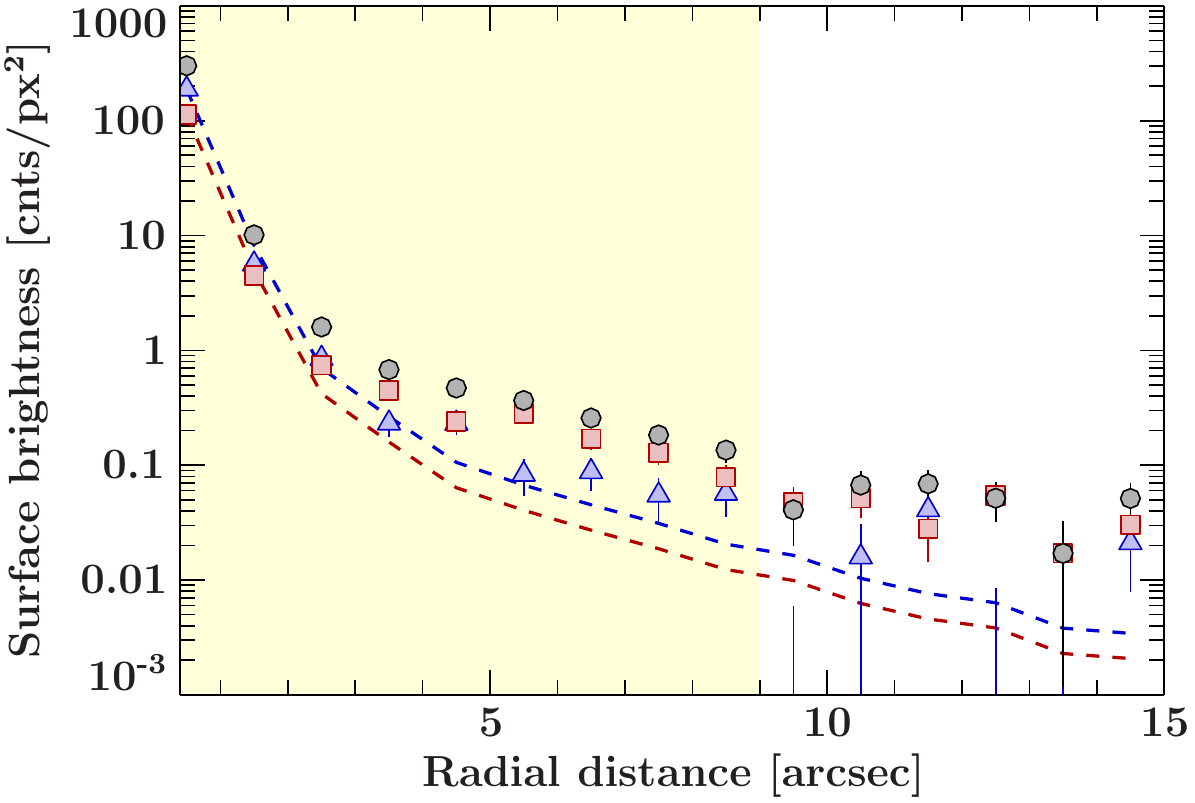}
  \caption{\textit{Left:} Difference map between the hard (1.5--10\,keV) and soft
    (0.3--1.5\,keV) band including counts of Ch~1, Ch~2, and Ch~3. Positive pixels (blue)
    depict a hard excess, negative pixels (red) a soft excess. Overlayed contours in black
    show H$_2$ 1--0 S(1) emission \citep{Maccagni2016}. \textit{Right:} Surface brightness
    profiles for the stacked map for the energy intervals 0.5--10\,keV (black/circles),
    0.3--1.5\,keV (red/squares) and 1.5--10\,keV (blue/triangles). The blue and red dashed
    lines show the \chandra\ PSF normalized to the hard (1.5--10\,keV) and soft
    (0.3--1.5\,keV) X-ray profile. The yellow shaded section marks the extent of the map
    on the left.}
  \label{fig:chandra_spat}
\end{figure*}
In order to quantify the extended and non-variable X-ray emission that
has been detected by \citet{Siemiginowska2016} based on the
\chandra\ observation Ch~1, we study a stacked image consisting of
Ch~1 and the more recent observations Ch~2 and Ch~3. The stacking was
performed using the standard \texttt{CIAO} task
\texttt{merge\_obs}. Figure~\ref{fig:chandra_spat} (left and right)
shows that the bulk of the photons and in particular hard X-rays above
$\sim$1.5\,keV are emitted from the unresolved core region with an
excess of $\sim$60\% above the soft X-rays within a radius of
3\arcsec. The PSF is encircling  99\% of the point-source flux
within $\sim$3\arcsec.  In contrast, soft ($\sim$0.3--1.5\,keV) X-ray
emission exceeds that in the hard band by $\sim$54\% in the extended
region between $\sim$3\arcsec\ and $\sim$8\arcsec.

\begin{figure}
  \resizebox{\hsize}{!}{\includegraphics{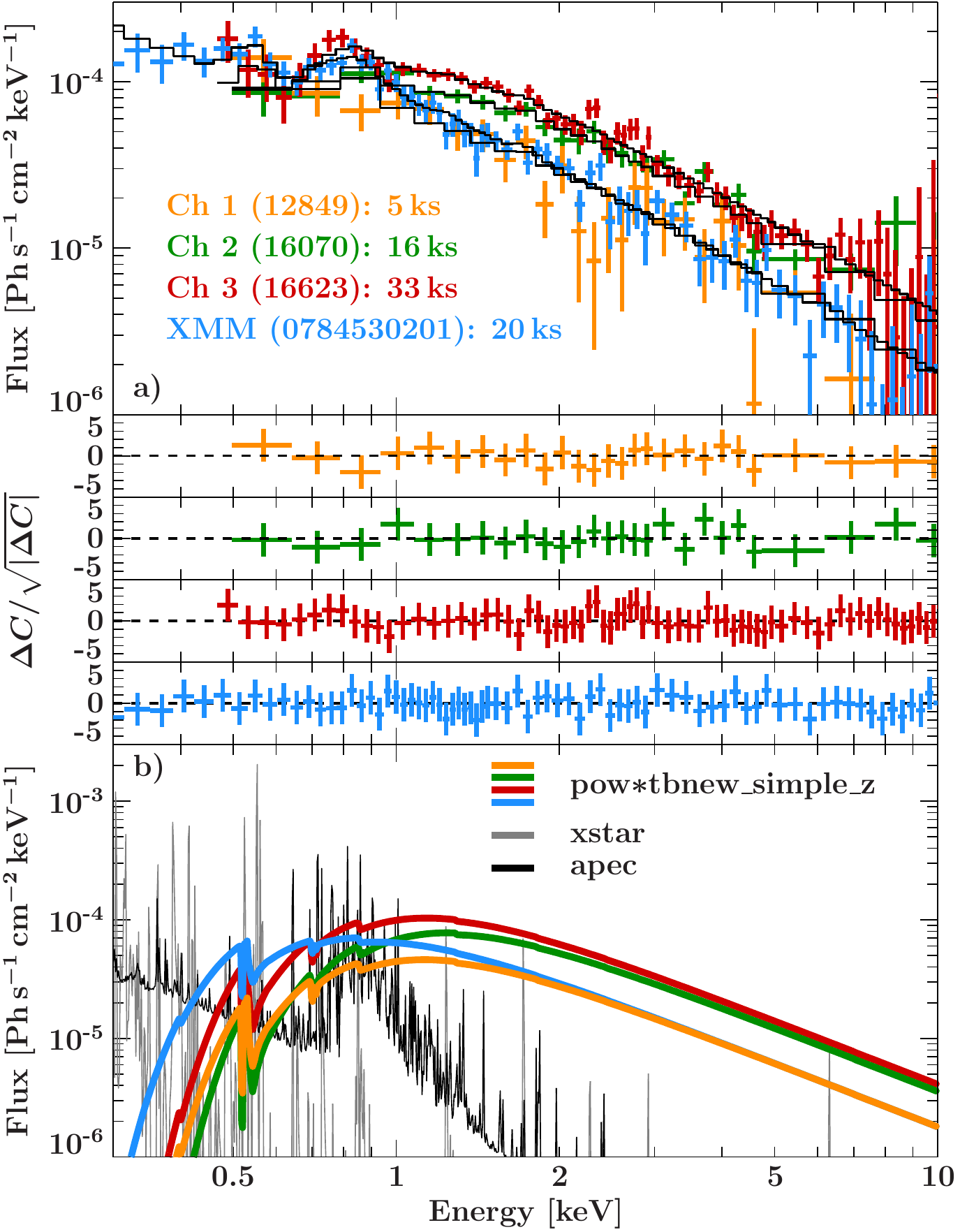}}
  \caption{\textit{Panel a:} Unfolded spectra for the observations
    Ch~1, 2, 3 and XMM in color with the individual model evaluations
    in black on top. The residuals are shown below. \textit{Panel b:}
    Model evaluations of the variable absorbed power law in color and
    the non-variable emission plasmas in gray (\texttt{xstar}) and
    black (\texttt{apec}).}
  \label{fig:xmm_spec}
\end{figure}

Despite the unprecedented spatial resolution of \chandra, the
soft-X-ray effective area of \xmm\ makes up twice and ten-times the
area of \chandra/ACIS at around 1\,keV and 0.5\,keV, respectively. We
therefore fit the \chandra\ spectra combined with recently acquired
\xmm\ data to unveil the origin of the extended emission (see
Fig.~\ref{fig:xmm_spec}). The hard-X-ray data follow a common
power-law of constant photon index ($\Gamma = 1.78$) that is absorbed
towards lower energies. Given the limited number of counts, we are not
sensitive to the continuum signatures of a possible ionized absorber
and instead apply a model for neutral absorbing gas
(\texttt{tbabs}). Above this continuum, \xmm\ allows us to confirm an
emission feature consistent with \ion{O}{vii} around 0.56--0.57\,keV
and a broad emission complex around 0.7--0.9\,keV, likely due to Ne
and Fe ($\sim 0.5-0.6$\,keV and $\sim 0.6-1.2$\,keV). The former is
described with the photoionized plasma component \texttt{xstar}
\citep[][]{Bautista2000,Porquet2000,Bautista2001} for a nuclear
irradiating power-law as measured, and an ambient gas density of
$10^{4}$cm$^{-3}$ \citep{Tingay2015}. We find the ionization parameter
to be $\log \xi = 0.04^{+1.13}_{-0.05}$. The latter broad feature is
best described with emission of a collisionally ionized plasma
($kT=0.75^{+0.07}_{-0.08}$\,keV) using \texttt{apec}
\citep{Smith2001}. This component dominates the soft X-rays and must
therefore account for a significant portion of the extended emission
in Fig.~\ref{fig:chandra_spat}. The low source flux prevents a
detailed line-diagnostic study with \xmm/RGS.

Besides the non-variable emission components \texttt{xstar} and \texttt{apec}, the only
parameters that turn out to be variable (on time scales of years) are the source-intrinsic
column density ($N_\mathrm{H} = 0.3-0.7 \times 10^{22}$\,cm$^{-2}$) and flux of the
incident hard X-ray power law ($5.4-11.6 \times 10^{-13}$\,erg\,cm$^{-2}$\,s$^{-1}$). We
use Cash statistics due to the low number of counts in Ch~1 and Ch~2. The simultaneous fit
describes all four data sets well with $\mathrm{C~(dof)}=444~(439)$. The resulting
parameters and uncertainties are listed in Table~\ref{tab:pars}.
\begin{table}
  \caption{Best-fit parameters for the simultaneous fit of the data
    from Table~\ref{tab:obs}.}
\label{tab:pars}
  \centering
  \scriptsize
  \resizebox{\columnwidth}{!}{
  \begin{tabular}[ht]{llllllllllllll}
    \hline\hline 
    abbrv. & \texttt{pow} & & \texttt{tbabs} & flux (0.5--10\,keV)  \\
    & norm & $\Gamma$ & \nh\ [$\times 10^{22}\,\mathrm{cm}^{-2}$] & [$\times 10^{-13}\,\mathrm{erg}\,\mathrm{cm}^{-2}\,\mathrm{s}^{-1}$] \\
    \hline
    Ch~1 & $\left(1.1^{+0.3}_{-0.2}\right)\times10^{-4}$ & $1.78^{+0.10}_{-0.09}$ & $0.5^{+0.3}_{-0.2}$ & 5.4 \\
    Ch~2 & $\left(2.2\pm 0.3\right)\times10^{-4}$ & `` & $0.69^{+0.18}_{-0.14}$ & 9.9 \\
    Ch~3 & $\left(2.5\pm 0.3\right)\times10^{-4}$ & `` & $0.55^{+0.14}_{-0.11}$ & 11.6 \\
    XMM  & $\left(1.09^{+0.15}_{-0.11}\right)\times10^{-4}$ & `` & $0.31^{+0.14}_{-0.09}$ & 5.8\\
    \hline
  \end{tabular}
  } \tablefoot{All spectra are fitted with \texttt{apec} (norm
    $2.0^{+0.6}_{-0.5} \times10^{-5}$; $kT=0.75^{+0.07}_{-0.08}$\,keV)
    and \texttt{xstar} ($N_\mathrm{H}=1\times 10^{25}\,$cm$^{-2}$
    based on an ambient gas density of $10^{4}\,\mathrm{cm}^{-3}$ and
    an assumed path length of 10\,pc; see \citet{Tingay2015}; $\log
    \xi = 0.04^{+1.13}_{-0.05}$). The unit of the \texttt{apec}
    normalization is
    $\left[10^{-14}/4\pi\,\left(D_\mathrm{A}\,(1+z)\right)^{2}\right]\,\int
    n_\mathrm{e}n_\mathrm{H}\,dV$ (see the \texttt{apec} manual for
    details). Overall, we apply a redshift of $z=0.0144$
    \citep{Doyle2005} and a Galactic column density of $5.7\times
    10^{19}\,$cm$^{-2}$ \citep{Kalberla2005}.}
\end{table}

\section{Discussion}
\label{sec:discussion}
We are able to report, for the first time in the literature, on
variable X-ray absorption and on the physics of the extended X-ray
emission in the CSO PKS~1718$-$649 that is for the most part unrelated
to nuclear feedback. We stacked all archival \chandra\ data available
for this object, and analyzed the combined spectra of \chandra\ and
\xmm. Our results form a two-fold picture. The bulk of the unresolved
core emission comprises Comptonized hard X-rays. The X-ray source may
be a corona close to the accretion disk \citep[e.g.,][and references
  therein]{Dove1997b}, the parsec-scale radio jet, or its jet base
\citep[e.g.,][]{Markoff2005,Dauser2013}. The spatially resolved and
extended environment beyond $\sim$600\,pc from the core, however,
predominantly emits soft X-rays.

Our proposed \xmm\ observation provides a sufficient amount of counts
below 1\,keV to investigate the nature of this extended and soft
emission. We detect two plasma emission components that are
non-variable on the probed time-scales of years: One is photoionized
by the incident continuum up to a level of $\log \xi =
0.04^{+1.13}_{-0.05}$, and the second dominates the soft emission and is
best described as a collisionally ionized and hot gas
($kT=0.75^{+0.07}_{-0.08}$\,keV or $T=(7.8-9.5)\times 10^{6}\,$K). A
significant contribution of reflected photons off the accretion disk
or surrounding matter can be tentatively ruled out: related codes
\citep[e.g., \texttt{xillver},][]{Garcia2013} predict strong
signatures of Fe~K\,$\alpha$ at $\sim$6.4\,keV as well as Mg, Si, and
S, which remains in contrast to our observations of highly ionized O and
Ne/Fe at $\lesssim 1\,$keV. Additional contributions to the soft
X-rays can likely be neglected. First, scattered radiation off the
extended gaseous and dusty environment is not required in our
fits. Second, the luminosity of a population of X-ray binaries would
lie roughly two orders of magnitude below the measured 0.5--8\,keV
luminosity given the weak star formation rate (SFR) of PKS~1718$-$649
\citep{Willett2010,Mineo2012}.

We now discuss the origin of these two components of gas. In the
optical, \citet{Filippenko1985} cross-identified the low-luminosity
AGN PKS~1718$-$649 with a low-ionization nuclear emission-line region
(LINER). In particular, the detection of [\ion{O}{iii}] brings up the
question of whether this emission could originate co-spatial and in
pressure equilibrium with the photo-ionized gas component that we
model with \texttt{xstar} in PKS~1718$-$649
\citep[e.g.,][]{Bianchi2006,Chakravorty2009}. \citet{Filippenko1985}
measure an [\ion{O}{iii}] intensity ratio of $(\lambdaup 4959 +
\lambdaup 5007)/\lambdaup 4363 \sim 12$, which implies temperatures of
$\gtrsim 6.2 \times 10^{4}$\,K \citep{Proxauf2014}. At these
temperatures, collisional ionization is the dominant process and
likely related to the parsec-scale jet.  In the absence of jet-shocked
regions, however, nuclear irradiation is still influential. Our
\texttt{xstar} modeling implies a highly photoionized plasma on sub-parsec
scales, traced by \ion{O}{vii}. At larger distances, the nuclear ionizing potential will
decrease and lead to a stratification of differently ionized species
of oxygen. While we are lacking spatial information on the
distribution of [\ion{O}{iii}] in PKS~1718$-$649, \citet{Ricci2015}
find [\ion{O}{iii}] to be distributed at around 100\,pc in a sample of
LINER galaxies, consistent with the size of a still expanding
Str\"omgen sphere in GPS sources \citep{Vink2006}. This photoionized
environment shifts to kiloparsecs in evolved AGN
\citep[e.g.,][]{RodriguezArdila2006}.

In PKS~1718$-$649, we spatially resolve a collisionally ionized gas
phase at $\lesssim$kpc distances. Based on its low overall luminosity
with $(2.5-5.4)\times 10^{41}\,$erg\,s$^{-1}$, we conclude that this
gas must be part of the extended and hot interstellar medium,
spatially distinct from the photoionized gas phase traced by
\ion{O}{vii}. In the CSO NGC~1052, the soft X-ray emission on kiloparsec
scales is well explained by jet excitation
\citep{Kadler2004a,Boeck2012}. This is most likely negligible in
PKS~1718$-$649, where the parsec-scale radio jet is too compact to drive
sufficiently energetic shocks into the interstellar medium (ISM) and to heat up this
gas. This is in line with the non-detection of coronal lines
\citep{Filippenko1985,RodriguezArdila2006} and the lack of notable
asymmetries of the X-ray emission on such large scales. We therefore
propose an alternative scenario, where supernovae (SNe) in the host
galaxy can feed the galactic halo with hot and X-ray-bright gas
\citep[see also, e.g.,][for similar conclusions based on a sample of
  star forming disk galaxies]{Strickland2004}.

Supernovae in the host galaxy are supported by the observation of active star
formation in PKS~1718$-$649 via H\,$\alpha$ and PAH (Polycyclic
Aromatic Hydrocarbon) emission
\citep{Kennicutt1983,Maccagni2014,Willett2010}. \citet{Sullivan2006}
study star-forming host galaxies of 100 confirmed SNe Ia. Their
results imply a SN rate of $\sim 3.2\times
10^{-2}\,\mathrm{galaxy}^{-1}\,\mathrm{yr}^{-1}$ for a galactic
stellar mass of $M_{\ast}\sim 4.9\times 10^{11}\,M_{\astrosun}$ for
PKS~1718$-$649 \citep{Maccagni2014}. When considering a SFR of $\sim
0.8-1.9\,M_{\astrosun}\,\mathrm{yr}^{-1}$ \citep[][using PAH
  signatures\footnote{PAH features at 6.2\,$\upmu$m
    ($EW=0.04\,\upmu$m) and 11.3\,$\upmu$m ($EW=0.13\,\upmu$m) are
    measured with \spitzer/IRS and a slit of
    11\arcsec$\times$168\arcsec.}]{Willett2010}, \citet{Sullivan2006}
infer a SN rate of $\sim 0.7-1.3\times
10^{-3}\,\mathrm{galaxy}^{-1}\,\mathrm{yr}^{-1}$. Very similar SN
rates, namely $(4\times 10^{-3}-4\times
10^{-2})\,\mathrm{galaxy}^{-1}\,\mathrm{yr}^{-1}$, can explain the
diffuse X-ray emission of M~81 \citep{Shelton1998,Page2003}, which is
classified as LINER as well \citep{Heckman1980}. Moreover, the gas
temperature that \citet{Page2003} determine for M~81 corresponds well
to that measured in our work.

An independent indicator for the presence of SNe is given by the
emission of different forms of hydrogen. While H$_2$ in the ISM is
generally too cold to emit, \citet{Maccagni2016} observe a disk-like
distribution in the inner few arcseconds ($\lesssim$kpc) of
PKS~1718$-$649. Roughly perpendicular to it, an outer disk of H$_2$ at
distances larger than 650\,pc aligns with neutral \ion{H}{i}
\citep{Maccagni2014} and H\,$\alpha$ \citep{Keel1991}. Among the
several H$_2$ excitation mechanisms at play \citep[e.g.,][and
  references therein]{Maloney1996,Rodriguez2004,Dors2012},
\citet{Maccagni2018} favor nuclear, non-thermal X-rays for the inner
few parsecs. Besides that, shock excitation by the parsec-scale jet or
nuclear UV radiation may also play a role in this compact environment. At
larger distances of hundreds to thousands of parsecs, H$_2$ appears
co-spatial with and most likely excited by the warm and diffuse soft
X-ray-emitting gas (Fig.~\ref{fig:chandra_spat}), which we suggest to
be due to the direct influence of SNe. The excitation can, however, to
some smaller degree also arise due to nuclear UV/X-ray emission or UV
photons of dense molecular star-forming regions
\citep{Puxley1990,Davies1998}.

\section{Conclusions}
\label{sec:conclusions}
In this letter, we investigated the nature of the extended X-ray
emitting gas in PKS~1718$-$649. By stacking all archival
\chandra\ data, we find this gas to primarily emit in soft X-rays. Our
recent observation by \xmm\ and its large effective area allow us to
perform a detailed spectral analysis of this emission. Besides a
photoionized ($\log \xi = 0.04^{+1.13}_{-0.05}$) gas phase on sub-parsec
scales, the bulk of the soft X-rays is emitted by diffuse, hot
($T=(7.8-9.5)\times 10^{6}\,$K), and collisionally ionized gas that
dominates the nuclear emission in the range $\sim 1-2.8\,$kpc. We argue that
supernovae are plausible candidates to power this region as opposed to
the overly compact parsec-scale jets of the young AGN. This conclusion is
driven from observations of active star formation in PKS~1718$-$649,
estimates on the expected SN rate, as well as the theoretically
predicted X-ray flux of SN remnants.

\begin{acknowledgements}  
We thank the internal referee at MPIfR, S.~Komossa, for her comments
that helped to significantly improve the clarity of this letter. We
are also grateful for fruitful discussions with S.~Toonen (on stellar
evolution and supernova progenitors), M.~Sasaki, P.~Zhou, and J.~Vink
(on supernova remnants), as well as N.~Przybilla (on atomic data and
[\ion{O}{iii}] diagnostics). T.B. acknowledges support by the
DLR grant DLR-50-OR-1606 and by the Dutch Research School for
Astronomy (NOVA). A.R.A thanks to CNPq for partial support to this
work. F.K. acknowledges funding from the European Union's Horizon 2020
research and innovation program under grant agreement
No. 653477. A.K. and M.L. acknowledge the support of the
Bundesministerium f\"{u}r Wirtschaft und Technologie (BMWi) through
Deutsches Zentrum f\"{u}r Luft- und Raumfahrt (DLR) grant
50OR1709. E.R. was partially supported by the Spanish MINECO grant
AYA2015-63939-C2-2-P and by the Generalitat Valenciana grant
PROMETEOII/2014/057. We thank J. E. Davis for the development of the
\texttt{slxfig} module that has been used to prepare the figures in
this work.  We made use of ISIS functions provided by ECAP/Remeis
observatory and MIT
(\url{http://www.sternwarte.uni-erlangen.de/isis/}) as well as the
NASA/IPAC Extragalactic Database (NED), which is operated by the Jet
Propulsion Laboratory, California Institute of Technology, under
contract with the National Aeronautics and Space Administration.
\end{acknowledgements}

\newcommand{\noop}[1]{}

\end{document}